\UseRawInputEncoding
\documentclass[letterpaper, 10 pt, conference]{ieeeconf}  

\IEEEoverridecommandlockouts                              

\usepackage{times}
\usepackage{epsfig}
\usepackage{graphicx}
\usepackage{amsmath}
\usepackage{amssymb}
\usepackage{comment}
\usepackage{footnote}
\usepackage{epstopdf}
\usepackage{comment}
\usepackage{multirow}
\usepackage{algorithm}
\usepackage{times}
\usepackage{epsfig}
\usepackage{graphicx}
\usepackage{amsmath}
\usepackage{amssymb}
\usepackage{wasysym}
\usepackage[ruled,vlined,algo2e]{algorithm2e}

\usepackage{color}
\usepackage{dsfont}	

\usepackage{adjustbox}

\usepackage{dblfloatfix}    
\usepackage{graphicx} 
\usepackage{authblk}
\usepackage{romannum}
\usepackage[table,xcdraw]{xcolor}
\usepackage{gensymb}

\title{\LARGE \bf
Stream-Based Ground Segmentation for Real-Time LiDAR Point Cloud Processing on FPGA}

\author[1]{Xiao Zhang, ~\IEEEmembership{Student Member,~IEEE, } 
        Zhanhong Huang,~\IEEEmembership{Student Member,~IEEE,} \\
        Garcia Gonzalez Antony,
        Witek Jachimczyk,
        and Xinming Huang,~\IEEEmembership{Senior Member,~IEEE}%

}

\begin{document}
\maketitle
\thispagestyle{empty}
\pagestyle{empty}

\begin{abstract}
This paper presents a novel and fast approach for ground plane segmentation in a LiDAR point cloud, specifically optimized for processing speed and hardware efficiency on FPGA hardware platforms. Our approach leverages a channel-based segmentation method with an advanced angular data repair technique and a cross-eight-way flood-fill algorithm. This innovative approach significantly reduces the number of iterations while ensuring the high accuracy of the segmented ground plane, which makes the stream-based hardware implementation possible.

To validate the proposed approach, we conducted extensive experiments on the SemanticKITTI dataset. We introduced a bird's-eye view (BEV) evaluation metric tailored for the area representation of LiDAR segmentation tasks. Our method demonstrated superior performance in terms of BEV areas when compared to the existing approaches. Moreover, we presented an optimized hardware architecture targeted on a Zynq-7000 FPGA, compatible with LiDARs of various channel densities, i.e., 32, 64, and 128 channels. Our FPGA implementation operating at 160 MHz significantly outperforms the traditional computing platforms, which is 12 to 25 times faster than the CPU-based solutions and up to 6 times faster than the GPU-based solution, in addition to the benefit of low power consumption.


\end{abstract}

 \footnote{This work was supported by The MathWorks. X. Zhang is with Department of Electrical and Computer Engineering, Worcester Polytechnic Institute, Massachusetts 01609, USA. (e-mail:xzhang25@wpi.edu).

W. Jachimczyk is with the Computer Vision and Autonomous Vehicles Team, The MathWorks, Massachusetts, 01760, USA (e-mail:wjachimc@mathworks.com).

Z. Huang, A. G. Gonzalez, and X. Huang are with the Department of Electrical and Computer Engineering, Worcester Polytechnic Institute, Massachusetts 01609, USA.  (e-mail:\{zhuang5, agarcia3, xhuang\}@wpi.edu). 
}


\section{Introduction}

LiDAR point cloud processing is important to the perception system of self-driving cars, robotics, and infrastructure surveillance. Unlike traditional camera-based sensors, LiDAR offers consistent performance across diverse lighting conditions, capturing detailed depth and shape information of the surroundings through Time-of-Flight (ToF) measurement.
\begin{figure}[h]
    \includegraphics[width=\linewidth]{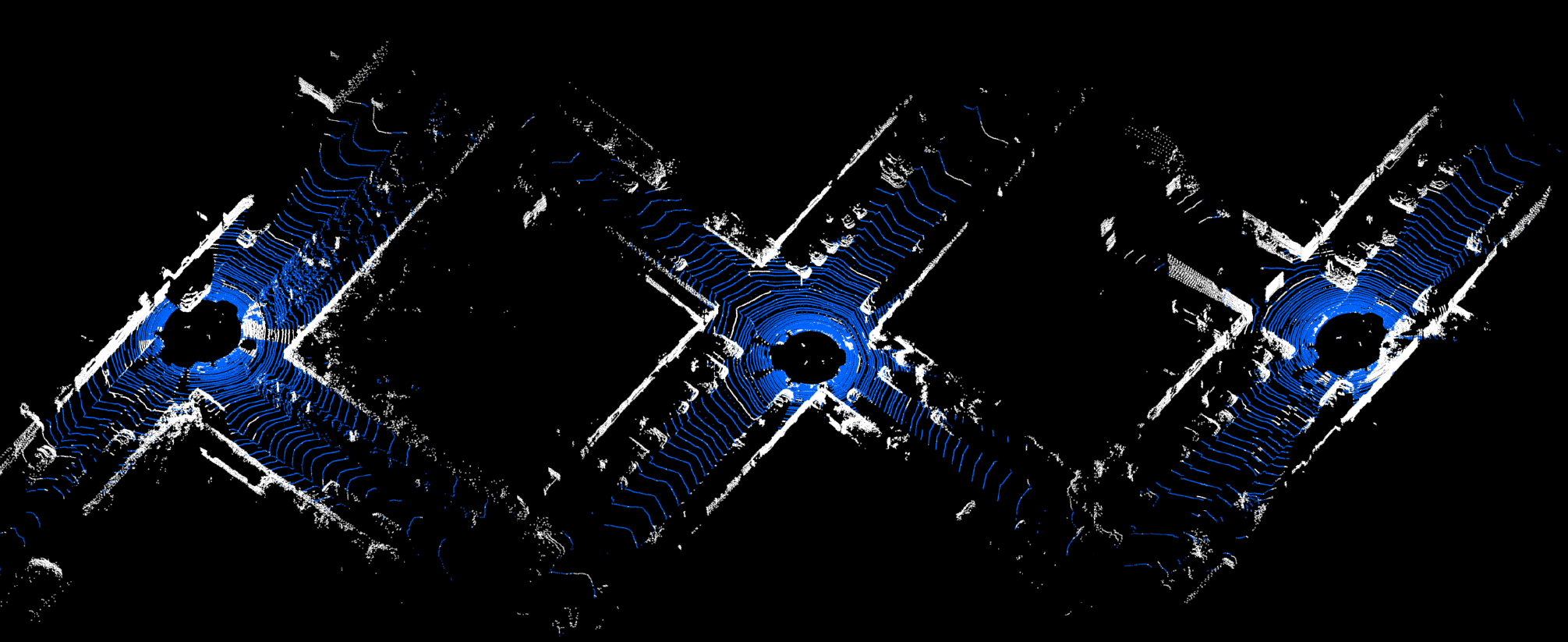}
    \caption{Ground Segmentation Results of T-Junction and Crossway Scenarios from the SemanticKITTI Dataset}
    \vspace{-0.4cm}
    \label{fig:ISICAS24_intro_demo}
\end{figure}

Traditional image sensors stand out for their high resolution and cost-effectiveness, but recent advancements in LiDAR technology have significantly improved point density while remaining cost-competitive. The evolution is happening from the HDL-32 E LiDAR sensor, with a 360$\degree$(H)x41.3$\degree$(V) field of view (FOV), generating 695,000 points per second, to the recent Hesai AT512, featuring a 120$\degree$x25.6$\degree$ FOV and producing 12,300,000 points per second. Additionally, devices like the Livox Mid-360 LiDAR, which provides a 360$\degree$x59$\degree$ FOV at approximately \$749, have become as economically viable as premium camera sensors, showcasing the rapid advancements and increasing accessibility of LiDAR technology. The expansion in applications of LiDAR point clouds is equally impressive, now encompassing a wide array of functions from detecting drivable areas \cite{lyuChipNetRealTimeLiDAR2019}, semantic \cite{xieRealTimeLiDARPoint2022} and panoptic segmentation\cite{Zhao_2021_ICCV}, to object detection and tracking \cite{asvadiDetectionTrackingMoving2015}, demonstrating the versatile utility of LiDAR technology in contemporary research and industry domains.

This work focuses on LiDAR point ground segmentation, a crucial step in the point cloud processing workflow that underpins significant perception tasks. Ground segmentation identifies drivable areas for autonomous vehicles and robots. Ground removal is also foundational for higher-level tasks such as object detection, tracking\cite{asvadiDetectionTrackingMoving2015}, and SLAM\cite{sunEnergyEfficientStreamBasedFPGA2023}, in which ground points are often removed first in the initial step of the algorithms.

Traditional point ground segmentation methods have been extensively explored, from model-based methods such as RANSAC \cite{fischlerRandomSampleConsensus1981} and grid-based strategies like 2.5D elevation maps \cite{douillardHybridElevationMaps2010} to range-image and channel-based approaches like depth ground segmentation \cite{bogoslavskyi2017efficient}. Many researchers advocate for the accuracy and semantic benefits of end-to-end learning-based solutions \cite{milioto2019rangenet++}\cite{zhou2020cylinder3d}\cite{xuRPVNetDeepEfficient2021}. However, ground segmentation, as a fundamental requirement across most LiDAR point cloud applications, faces significant challenges of variations in sensor configuration, which include differences in channel numbers (resolution), fields of view (FOV), and laser scanning mechanisms, ranging from mechanical spinning LiDAR to semi-solid-state and fully solid-state LiDAR systems. Such variability of data sources can impact the accuracy of learning-based ground segmentation models.


Deploying real-time processing on edge platforms posts additional challenges. Existing LiDAR point cloud ground segmentation methods \cite{bogoslavskyi2017efficient} \cite{velasCNNVeryFast2018} can achieve execution times within 100 ms for 64-channel sensors, which satisfies real-time criteria with current point density. However, as previously mentioned, the computational burden escalates considerably with higher-resolution sensors. As the size of the input point cloud increases over tenfold in the latest LiDAR sensors, maintaining real-time streaming workflows becomes particularly challenging when ground segmentation is merely one component of the pipeline. Furthermore, the high power consumption of a high-end GPU, which the learning-based models require, becomes an important concern in real-time edge applications. In this context, FPGA-based accelerator designs have been shown to offer a power-efficient solution for real-time point cloud processing. \cite{lyuChipNetRealTimeLiDAR2019}\cite{sunEnergyEfficientStreamBasedFPGA2023}\cite{zhangRealTimeFastChannel2022}.

In response to these challenges, this paper presents a fast, hardware-friendly, channel-based ground segmentation algorithm and its implementation on an FPGA platform. Our architecture offers a substantially lower processing time compared to existing methods while maintaining high accuracy and low power consumption. The design also supports LiDAR sensors with various resolutions. The contributions of this work can be summarized as follows:

\textbf{Channel-based LiDAR Ground Segmentation with Angular Data Repair and Cross-Eight-Way Flood-Fill.} We introduced a hardware-efficient channel-based method for LiDAR point cloud ground plane segmentation. Our approach includes streamlined angular data repair techniques for enhanced efficiency. To overcome the challenges of iterative processing in hardware, we developed and validated a cross-eight-way flood-fill algorithm, ensuring high performance while limiting the iteration requirement.

\textbf{SemanticKITTI Dataset Evaluation with BEV-Based IoU Metric for LiDAR Segmentation.} Comprehensive experiments utilizing the SemanticKITTI dataset were performed to assess our ground segmentation method. We introduced a bird's-eye view (BEV)-based intersection-over-union (IoU) metric, designed explicitly for effectively evaluating the LiDAR segmentation from an area perspective. In-depth discussions of various metrics are presented through a study using SemanticKITTI benchmarks. Post-publication, the code will be made available on GitHub 

\textbf{Optimized FPGA Deployment for LiDAR Segmentation with Speed Acceleration and Low Power Consumption} We designed an optimized hardware architecture to deploy our channel-based ground segmentation algorithm on the Zynq-7000 FPGA, achieving compatibility with 32, 64, and 128 channel LiDAR sensors. Operating at 160MHz, our FPGA implementation significantly outperforms traditional computing platforms, achieving up to 25 times speedup over state-of-the-art (SOTA) CPU-based implementations and 6 times speedup over GPU-based implementations, in addition to the benefit of much lower power consumption. 



\section{Related Work}
Ground segmentation from LiDAR point cloud data is essential for many applications, especially in autonomous driving, where precise environmental perception underpins navigation and driving safety. The existing literature reveals a diverse array of ground segmentation techniques, each offering distinct advantages and facing unique challenges. This section systematically reviews them in four main categories: ground modeling methods, grid-based methods, range image and channel-based methods, and learning-based methods. 

\subsection{Ground Modeling Methods} Ground modeling represents an intuitive approach to segmenting ground points, assuming that ground points can be described through specific geometric or statistical models. These algorithms often need to traverse the unorganized points in 3D space. A notable approach, RANdom SAmple Consensus (RANSAC) \cite{fischlerRandomSampleConsensus1981}, introduced by Fischler and Bolles, estimated a 2D geometric plane by determining unknown parameters through points within an orthogonal threshold. Additionally, a zone-based, region-wise ground segmentation method \cite{limPatchworkConcentricZoneBased2021} utilized Principal Component Analysis (PCA) to address false positives and computational intensity. Beyond planar models, Gaussian-based methods \cite{chenGaussianProcessBasedRealTimeGround2014} and Markov Random Field (MRF) based methods \cite{FastPointCloud} offered alternative modeling strategies. However, these methods often fell short in accurately identifying realistic ground areas, particularly in complex terrains with steep slopes or curbs, and tended to be computationally intensive.

\subsection{Grid-based Methods} Grid-based methods aim to mitigate computational demands and inefficiencies stemming from the LiDAR point cloud's uneven distribution in sparse 3D space. The concept of a 2D occupancy grid map, introduced by Moravec and Elfes \cite{moravecHighResolutionMaps1985}, converted the noisy 3D point cloud into a 2D grid-based bird's-eye view (BEV) occupancy map. Building on this, the elevation map approach encoded the point cloud into a 2.5D grid, with height represented by relative \cite{thrunStanleyRobotThat2006} or mean height of points \cite{asvadiDetectionTrackingMoving2015}. Techniques such as gradient classification and clustering then enabled the differentiation between ground and objects \cite{douillardHybridElevationMaps2010} \cite{mengTerrainDescriptionMethod2018}. While BEV 2D space and grid down-sampling substantially reduced computational costs, challenges remained in mislabeling low-height objects and inaccuracies due to height variance within a grid cell.

\begin{figure*} [h]
    \begin{center}
        \includegraphics[width=\linewidth]
        {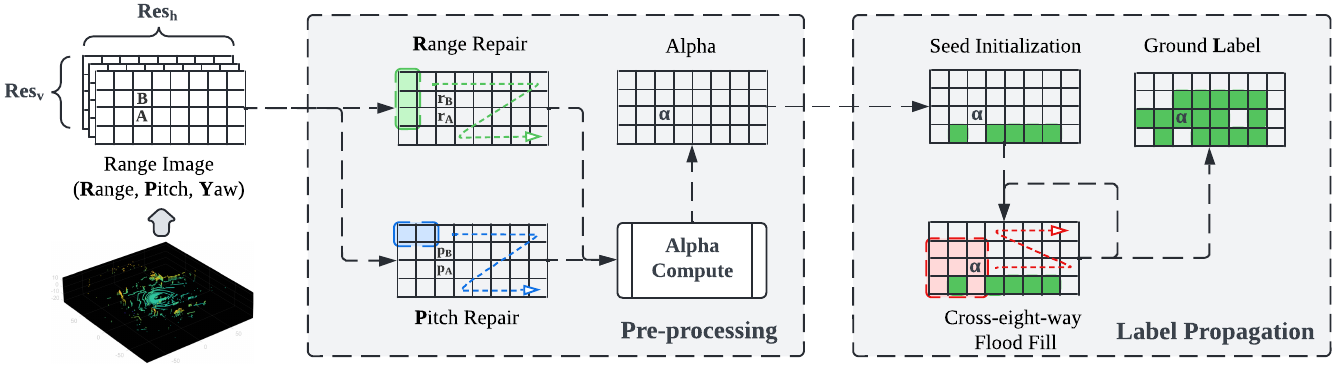}
        \caption{Pipeline of Channel-based Ground segmentation: This diagram illustrates the process from extracting range images from organized point clouds to segmentation, with frame sizes determined by the LiDAR sensor's vertical and horizontal resolutions (e.g., 64x2048 for OS-64 LiDAR)}
        \label{fig:ISICAS24_alg_pipeline}
    \end{center}
\end{figure*}

\subsection{Range Image and Channel-based Methods} Range image offers an alternative for mapping the 3D point cloud to 2D space through spherical projection, aligned with the mechanical LiDAR sensor's scanning mechanism. Channel-based methods process points by LiDAR scan channels, with different techniques for intra- and inter-channel processing. For instance, \cite{narksriSloperobustCascadedGround2018} used the Cartesian distance between consecutive channels to filter non-ground points and applied RANSAC for multi-region plane fitting.\cite{bogoslavskyi2017efficient} leveraged pitch angle differences derived from polar coordinates to identify ground points, followed by seed initialization and propagation. Another method proposed by Chu et al. \cite{chuFastGroundSegmentation2017} utilized both angular and distance features as thresholds. These methods enhanced point cloud traversal efficiency but typically required extensive iterations for optimal performance.

\subsection{Learning-based Methods} Adopting deep learning brought learning-based methods to the research frontier. GndNet\cite{paigwarGnd2020}, for example, estimated the ground plane using a grid-based representation, with segmentation results determined by ground elevation thresholds. Other networks, such as RangeNet++\cite{miliotoRangeNetFastAccurate2019}, Cylinder3D\cite{zhou2020cylinder3d}, and RPVNet\cite{xuRPVNetDeepEfficient2021}, offered end-to-end solutions for semantic segmentation, which included ground labeling. While deep learning approaches excelled in dataset evaluations, they demanded substantial GPU resources with large power consumption. Their performance also depended on the similarity between the actual input data and the training data scenarios and sensor configurations.

\subsection{Hardware Acceleration Implementations}
In this subsection, we delve into hardware acceleration designs pertinent to LiDAR point cloud ground segmentation, with an emphasis on classical and learning-based approaches. 

 Classic methods have been explored to enhance RANSAC processing speed on the FPGA platform. Tang et al. \cite{tangFPGAImplementationRANSAC2013} and Vourvoulakis et al. \cite{vourvoulakisAccelerationRANSACAlgorithm2016} have proposed acceleration techniques specifically for image-based RANSAC algorithms. Additionally, Zhou et al. \cite{zhouFPGAbased3D2DRANSAC2020} developed an FPGA framework aimed at expediting feature point matching for RGB-D images. These initiatives predominantly target 2D image processing rather than direct point cloud manipulation. For GPU acceleration, Baker and Sadowski \cite{bakerGPUAssistedProcessing2013a} constructed a GPU-assisted ground segmentation system within the Robotic Operation System (ROS), specifically tailored for the HDL-64E LiDAR Sensor.

Among the learning-based approaches, Lyu et al. introduced ChipNet \cite{lyuChipNetRealTimeLiDAR2019} as a real-time FPGA-based solution for segmenting drivable regions from LiDAR data. Distinguished by its acceleration of both 3D and 2D convolution processes, ChipNet employs a CNN architecture tailored for high-speed operations. Further progress has been observed in accelerating learning models for semantic segmentation. Research by Xie et al. \cite{xieRealTimeLiDARPoint2022}, Jia et al. \cite{jiaDesignImplementationRealtime2021}, and Vogel et al. \cite{vogelEfficientAccelerationCNNs2019} introduced architectures designed for semantic segmentation, highlighting an ongoing shift towards improving the processing speed of deep learning models on hardware platforms.




\section{Algorithm Design}
The elevation angle-based ground segmentation method is based on three essential assumptions:

(I) A range image can be projected from each frame of the LiDAR point cloud data.

(II) The bottom edge of the range image should align with the ground. The observation of the ground starts from the lower channels. 

(III) The curvature of the ground plane is limited.

The pipeline of our proposed channel-based ground segmentation is shown in Figure. \ref{fig:ISICAS24_alg_pipeline}. To adhere to the first assumption, we apply spherical projection. This operation transforms an unorganized, raw point cloud into an organized format. The range image also facilitates efficient indexing of the sparse 3D point cloud using a dense 2D graph, significantly reducing the algorithm complexity.

During the pre-processing phase, we calculate the pitch angle difference matrix $Alpha$ to represent channel-wise vertical curvature on the range image and implement repair techniques to mitigate the impact of missing points.

The proposed range image based segmentation method is inspired by classic image processing. Predicated on the latter two assumptions, the channel-wise label propagation unfolds in two main stages: bottom-up labeling initialization and cross-eight- way flood-fill based on the $Alpha$ matrix. 

Detailed discussions, including algorithm's design, experimentation, and performance analysis, will be presented in this section.

\begin{figure}[H]
    \includegraphics[width=\linewidth]{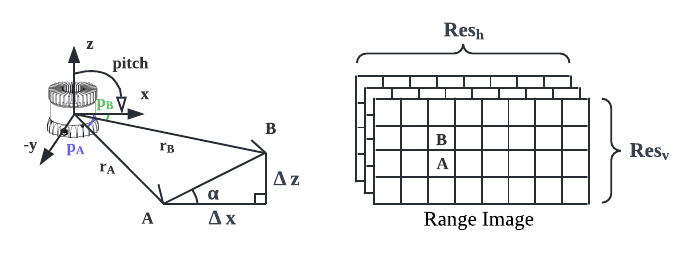}
    \caption{Illustration of $Alpha$ Calculation: This figure demonstrates the calculation of $\alpha$ between successive points $A$ and $B$ from adjacent laser channels in a LiDAR scan, both positioned in the same column of the range image. The ranges of $A$ and $B$ are denoted as $r_A$ and $r_B$, while their pitch angles are represented by $p_A$ and $p_B$, respectively}
    \vspace{-0.4cm}
    \label{fig:ISICAS24_alg_demo}
\end{figure}

\subsection{Pre-processing}

During pre-processing, we compute an angle matrix, $Alpha$, to characterize the curvature between adjacent LiDAR channels. As shown in Figure. \ref{fig:ISICAS24_alg_demo}, we calculate the pitch angle difference $\alpha$ between vertically adjacent points in successive vertical channels of the range image, exemplified by points $A$ and $B$. The pitch angle difference is derived from equations as follows:

\begin{equation}
    \alpha = \arctan2 (\Delta Z, \Delta X)
\end{equation}

Where the $\Delta X$ and $\Delta Z$  are calculated by:
\begin{equation}
\begin{split}
    \Delta X = | r_a \sin{p_a} - r_b \sin{p_b} | \\ 
    \Delta Z = | r_a \cos{p_a} - r_b \cos{p_b} |
\end{split}
\end{equation}

As illustrated as the pre-processing module in Figure \ref{fig:ISICAS24_alg_pipeline}, during the formation of the $Alpha$ matrix, each $\alpha$ value is assigned to the position corresponding to the lower channel point, exemplified by point $A$. To maintain a consistent matrix size throughout the pipeline, the top row of the $Alpha$ is populated with the values from the second row.

\begin{figure}[H]
    \includegraphics[width=\linewidth]{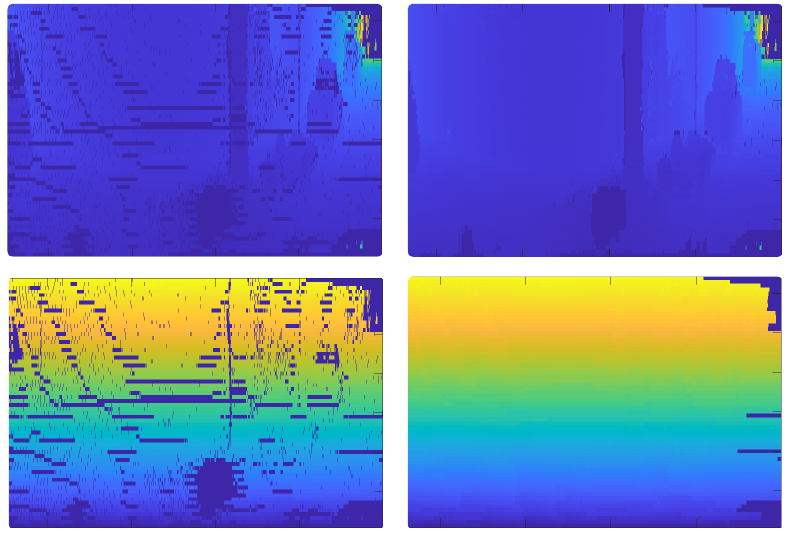}
    \caption{Data Repair on Range and Pitch Frame: \textbf{Top Left} Range frame before repair. \textbf{Top Right} Range frame after repair. \textbf{Bottom Left} Pitch frame before repair. \textbf{Bottom Right} Pitch frame after repair}
    \vspace{-0.4cm}
    \label{fig:ISICAS24_alg_smooth}
\end{figure}

In this work, we introduce a refined frame value repair technique to address data quality issues, such as missing values and outliers in the raw point cloud data or arising from spherical projection. Diverging from conventional methods that apply convolution indiscriminately across the entire $Alpha$ matrix, our approach employs a two-pronged strategy: column-wise average repairing on the range matrix, $R$, coupled with row-wise nearest neighbor correction on the pitch matrix, $P$. This dual strategy not only facilitates more effective smoothing of the $Alpha$ matrix but also helps boost hardware performance by enabling parallel processing of multiple matrices from consecutive frames. 

The left side of Figure. \ref{fig:ISICAS24_alg_smooth} illustrates the critical challenges of missing points and outliers affecting inter-channel computations. By analysis of statistical features within frames $R$ and $P$, we have developed targeted repairing strategies: a column-wise average repairing for the range frame $R$ and a row-wise data nearest neighbor repairing for the pitch frame $P$.

\subsubsection{\textbf{Range Repair}} 

The range frame $R$ encapsulates precise depth information, which is crucial for depicting an object's shape and location as well as for ground detection. To address missing points and outliers without altering the depth distribution's local features, we introduce an average repair algorithm. 

The average repair process employs an average filter along the range frame, specifically using a column-wise window to correct invalid points in the $R$ frame, as illustrated in Figure \ref{fig:ISICAS24_alg_pipeline}. This process involves comparing the $\alpha$ differences between pairs of points equidistant from the window's center $r_c$. If the range difference $|r_{c+s_{up}} - r_{c-s_{down}}|$ for any pair is less than $repairRangeThresh$, the center invalid value is repaired with the average range value derived from all valid point pairs within the window.

In our implementation, as illustrated in the upper right of Figure \ref{fig:ISICAS24_alg_smooth}, the range repair step is configured with a step size of 2 and a window size of 5. The range repairing effectively restores the integrity of the range frame $R$ and preserves critical edge features essential for plane detection.

\subsubsection{\textbf{Pitch Repair}}
The pitch frame $P$ exhibits a clear uniform distribution, attributed to the LiDAR sensor's fixed laser channel angle. Here, we employ a nearest-neighbor value-repairing approach tailored for rows. As the scan progresses, we update the nearest neighbor buffer with valid column values and replace invalid ones with the buffer's current value. The bottom right of Figure \ref{fig:ISICAS24_alg_smooth} shows the repaired frame, highlighting a distinct vertical angular distribution of the LiDAR channels.

\subsection{Label Propagation}
The label propagation phase encompasses label initialization and iterative expansion. It begins with identifying the first valid ground point from the bottom channel, using a predetermined threshold to establish the initial label seed. Subsequently, we introduce an enhanced flood-fill technique for label expansion, which significantly reduces iterations without compromising performance. This process is illustrated as the incremental labeling module in Figure \ref{fig:ISICAS24_alg_pipeline}.
\subsubsection{Seed Initialization}

Based on the assumption (\Romannum{2}), ground points are expected to originate from the lower (bottom) channels situated at the bottom of the range image. Consequently, we initiate the process by establishing a seed for each column within the range image. For each column, the first valid point is identified. If the corresponding $\alpha$ value is less than or equal to $seedThresh$, this point is designated as the seed for the respective column.

\subsubsection{Cross Eight Way Flood-fill}

\begin{figure}[H]
    \includegraphics[width=\linewidth]{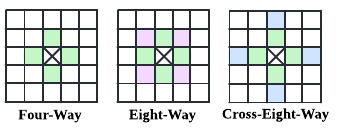}
    \caption{Demonstration of Flood-Fill Methods: The central point's one-step neighbors (s1) are marked in green, two-step neighbors (s2) in blue, and diagonal neighbors in purple.}
    \vspace{-0.4cm}
    \label{fig:ISICAS24_alg_floodfill}
\end{figure}

Utilizing seeds identified in the preceding phase, we employ a flood-fill technique to expand the ground label area, as detailed in Figure \ref{fig:ISICAS24_alg_floodfill}. Traditional variations, such as four-way neighbor of four (N4)  or eight-way flood-fill (N8), offer differing scopes of search. The four-way approach, limited by its narrow search area, often results in longer iteration times for optimal performance. While the eight-way method extends the search to include diagonal neighbors alongside orthogonal ones, its effectiveness in graphic applications does not directly translate to our context. Given the connectivity in the pitch angle, the elevation difference matrix $Alpha$ shows more robustness in the orthogonal way. 

To address these challenges, we introduce a cross-eight-way flood-fill approach. This method evaluates connectivity by comparing the value difference between a center point and its four immediate orthogonal neighbors $|\alpha_{c} - \alpha_{s1}|$ against a threshold $alphaThresh$. If the criterion is not met, it then examines the neighbors in the same direction $|\alpha_{s2} - \alpha_{s1}|$, and if consistent, further assesses $|\alpha_{c} - \alpha_{s2}|$. A center point is designated as ground if any neighbor satisfies the threshold criteria and is already labeled as ground.

\begin{figure}[H]
    \includegraphics[width=\linewidth]{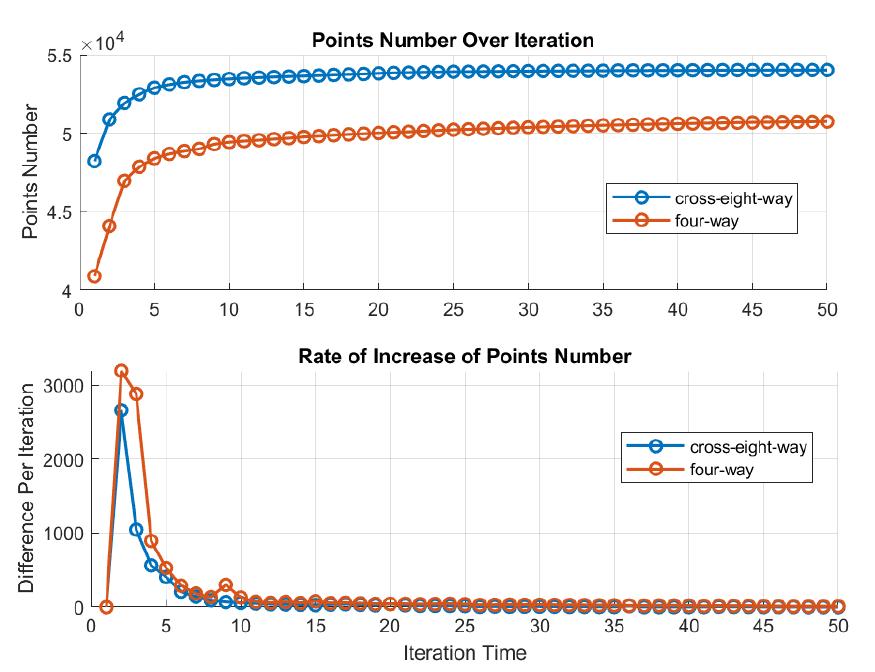}
    \caption{Four-way and Cross-eight-way Flood-fill Propagation (Top) and Point Number Increasing Rate (Bottom)}
    \vspace{-0.4cm}
    \label{fig:ISICAS24_alg_iternum}
\end{figure}

As demonstrated in Figure \ref{fig:ISICAS24_alg_iternum}, our proposed cross-eight-way flood-fill method exhibits superior efficiency to the traditional four-way flood-fill. This efficiency is evident from the significantly larger number of points filled after the first iteration, attributable to the cross-eight-way method's expanded search area. Moreover, the rate of increase in filled points diminishes more rapidly in subsequent iterations, indicating a quicker convergence toward optimal fill.

\subsection{Data Set and Metrics}

\subsubsection{Dataset}
SemanticKITTI \cite{behley2019semantickitti} is a widely recognized benchmark dataset for semantic scene understanding in the context of autonomous driving. Built upon the KITTI \cite{geiger2013vision}, it extends the original dataset with dense, point-level annotations for a comprehensive set of semantic categories across all lidar scans. SemanticKITTI offers an unparalleled resource for researchers and practitioners, providing over 43,000 scans across 22 sequences, where each point in the 3D point cloud is labeled with one of 28 semantic classes, including but not limited to vehicles, pedestrians, buildings, vegetation, and ground related areas. This rich annotation facilitates the development and evaluation of algorithms for tasks such as semantic segmentation, object detection, and scene prediction in the domain of lidar-based perception. In this work, we built a ground segmentation experiment method and introduced novel bird-eye-view evaluation metrics based on the SemmanticKITTI. We combine multiple class labels, including No.40 roads, No.44 parking, No.48 sidewalks, and No.49 other grounds, as the ground truth when evaluating the ground segmentation results. 


\subsubsection{Metrics}

Two predominant metrics commonly employed to assess the performance of semantic segmentation models are the Intersection over Union (IoU) and the F1 Score.

The F1 Score serves as the harmonic mean of $\text{Precision}$ and $\text{Recall}$, striking a balance between these two crucial aspects. This metric is especially valuable in contexts where the distribution of classes is uneven, as it effectively measures the model's accuracy in identifying true positives while minimizing the inclusion of irrelevant data points (false positives and false negatives). The F1 Score is mathematically represented as:

\begin{equation}
    \text{F1 Score} = 2 \times \frac{\text{Precision} \times \text{Recall}}{\text{Precision} + \text{Recall}}
\end{equation}

Where $\text{Precision}$ is defined as $\frac{TP}{TP + FP}$, with $\text{TP}$ being the number of true positives and $\text{FP}$ the number of false positives.
$\text{Recall}$ is defined as $\frac{TP}{TP + FN}$, with $\text{FN}$ the number of false negatives.

IoU quantifies the percentage overlap between the ground truth and the prediction from a segmentation model. It is calculated by dividing the size of the intersection of the predicted and ground truth masks by the size of their union. Mathematically, IoU is expressed as:

\begin{equation}
     \text{IoU} = \frac{\text{Area of Overlap}}{\text{Area of Union}} = \frac{\text{TP}}{\text{TP} + \text{FP} + \text{FN}}
\end{equation}

Traditionally, both metrics are calculated based on the point classification of the range image. Although the range image-based evaluations are suitable for object detection or multi-class semantic segmentation, they introduce inherent biases in ground plane evaluation from the area side. This is due to the perspective distortion inherent in spherical projections, which can disproportionately affect the assessment of ground plane segmentation. 

\begin{figure}[H]
    \includegraphics[width=\linewidth]{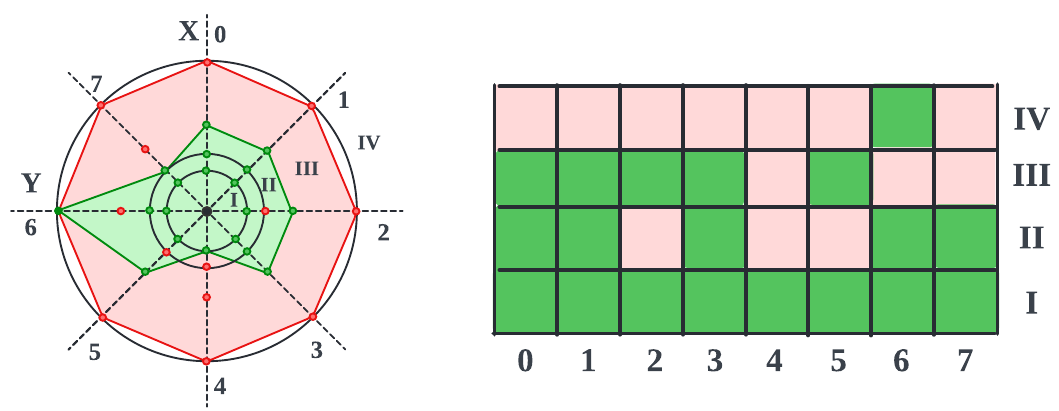}
    \caption{Ground Segmentation Representation from Bird-Eye-View and Range Image: This illustration showcases ground segmentation using a 4-channel LiDAR with a horizontal resolution of 45 degrees (left), corresponding to a 4-by-8 range image (right). Left and right figures evaluate true positives (green) and false negatives (red) from distinct perspectives.}
    \vspace{-0.4cm}
    \label{fig:ISICAS24_alg_metric}
\end{figure}

\begin{table*}[t]
\begin{center}
\label{tb:evaluation_semmantic}
\caption{Performance Evaluation of Ground Segmentation on SemmantickKITTI}
 \begin{adjustbox}{width=17cm, center}
\begin{tabular}{cc|ccc|ccc|ccc|ccc}
\hline \hline
\multicolumn{2}{c|}{\textbf{Method}} & \multicolumn{3}{c|}{\textbf{RANSAC}}                & \multicolumn{3}{c|}{\textbf{DepthGround \{100 Iter\}}}           & \multicolumn{3}{c|}{\textbf{DepthGround \{10 Iter\}}} & \multicolumn{3}{c}{\textbf{Ours \{10 Iter\}}}       \\ \hline 
\textbf{Sequence}  & \textbf{Frame}  & \textbf{F1-RI} & \textbf{IoU-RI} & \textbf{IoU-BEV} & \textbf{F1-RI} & \textbf{IoU-RI} & \textbf{IoU-BEV} & \textbf{F1-RI}  & \textbf{IoU-RI}  & \textbf{IoU-BEV} & \textbf{F1-RI} & \textbf{IoU-RI} & \textbf{IoU-BEV} \\ \hline 
00                 & 4541            & 85.93          & 76.65           & 64.04            & 87.48          & 78.51           & 70.09            & 78.69           & 65.62            & 48.80            & 86.66          & 77.29           & 70.55            \\ \hline
01                 & 1101            & 78.94          & 67.37           & 36.36            & 83.02          & 70.09           & 41.28            & 78.09           & 64.56            & 33.15            & 81.87          & 71.75           & 41.57            \\ \hline
02                 & 4661            & 88.83          & 80.33           & 62.50            & 88.50          & 79.76           & 73.46            & 80.22           & 67.42            & 57.95            & 86.23          & 76.20           & 70.73            \\ \hline
03                 & 801             & 74.74          & 61.68           & 40.68            & 83.47          & 72.44           & 50.33            & 76.14           & 62.03            & 41.64            & 78.74          & 66.02           & 44.13            \\ \hline
04                 & 271             & 81.50          & 69.11           & 52.60            & 90.02          & 81.89           & 70.10            & 87.04           & 77.15            & 62.30            & 88.32          & 79.30           & 67.74            \\ \hline
05                 & 2761            & 84.58          & 74.14           & 61.23            & 87.81          & 78.55           & 67.38            & 77.25           & 63.50            & 47.92            & 85.35          & 74.73           & 66.68            \\ \hline
\multicolumn{2}{c|}{Mean}            & 85.36          & 75.66           & 59.28            & 87.35          & 78.00           & 67.31            & 78.88           & 65.73            & 50.28            & 85.47          & 75.40           & 66.05            \\ \hline
\end{tabular}
\end{adjustbox}

\end{center}
\label{tab:ISICAS24_alg_performance}
All values presented in the table are expressed as percentages.
\end{table*}

The conventional evaluation metrics for semantic segmentation may not adequately address the unique property associated with ground plane segmentation. These challenges include the non-uniform density of points at varying distances from the sensor, the necessity to circumvent biases introduced by projection methods, and the aspect of preserving geometric consistency throughout the segmentation process. This discrepancy is illustrated in Figure \ref{fig:ISICAS24_alg_metric}. Assuming a vertically mounted LiDAR sensor scanning an open area, the segmentation algorithm categorizes points into ground (green) and non-ground (red) classifications.

The left image in Figure. \ref{fig:ISICAS24_alg_metric} offers a bird's-eye view of the detected ground plane, showcasing the areas identified as ground. Conversely, the right image displays how these classifications are represented in a range image, with a noticeable predominance of correctly detected ground points over non-ground ones. However, this representation masks a critical flaw: despite the apparent abundance of correctly identified ground points in the range image, the actual proportion of accurately detected ground area, as viewed from the bird's-eye perspective, falls below 50 percent. This discrepancy underscores the limitations of conventional evaluation metrics, which might not adequately reflect the actual effectiveness of ground segmentation algorithms, especially regarding spatial accuracy and distribution consistency. 

Recognizing the limitations of traditional point-based evaluation metrics for ground plane segmentation, particularly their inability to adequately address the unique challenges and projection-based biases inherent in this task, here we propose the Bird-Eye-View Intersection over Union (BEV-IoU) as a novel metric specifically designed to assess ground segmentation algorithms. This metric is crafted to directly address the issues of varying point densities along the distance and to meet the critical need for geometric consistency in segmented outputs.

\begin{equation}    
    \text{IoU-BEV} = \frac{ \text{Poly}_{gt} \cap \text{Poly}_{pred}}{\text{Poly}_{gt} \cup \text{Poly}_{pred}}
\end{equation}

Where the $\text{Poly}_{gt}$ denotes the polygon formed by the ground truth of ground points on BEV, $\text{Poly}_{pred}$ denotes the polygon formed by predicted ground points on BEV. To form the polygon as shown in the left image of Figure. \ref{fig:ISICAS24_alg_metric}, we first make some notion: 

$P = \{p_1, p_2, ..., p_n\}$ as the set of points in the 3D point cloud, where each point $p_i = (x_i, y_i, z_i)$ is a tuple representing the 3D coordinates of the $i$-th point.
$P_{2D} =\{p_1', p_2', ..., p_n'\}$ as the projection of $P$ onto the XY-plane, where each point $p_{2D,i} = (x_i, y_i)$.

$P_{polar} = \{(\theta_i, \rho_i) | \theta_i = \arctan2(y_i, x_i), \rho_i = \sqrt{x_i^2 + y_i^2}\}$ as the set of points in $P_{2D}$ converted to polar coordinates. Here, $\arctan2(y, x)$ computes the angle $\theta$ between the positive x-axis and the point $(x, y)$, and $\rho$ is the distance from the origin to the point.

$E = \{e_{\theta} | \theta = -180^\circ, ..., 180^\circ\}$ as the set of edge points in polar coordinates, where each $e_{\theta}$ is the point with the maximum $\rho$ within a 1-degree angular sector centered at $\theta$.

The \text{Poly} (Resolution with $1 ^\circ$) can be represented by the set $E$, where each edge point $e_{\theta}$ is determined as follows:

1.\textbf{ Angular Sector Identification}: For each degree $\theta$, identify the subset of points $S_{\theta} \subset P_{polar}$ such that each point $(\theta_i, \rho_i) \in S_{\theta}$ satisfies $\theta - 0.5^\circ \leq \theta_i < \theta + 0.5^\circ$. 

2. \textbf{Maximum Distance Selection}:
   - For each subset $S_{\theta}$, select the point with the maximum $\rho$, denoted as $e_{\theta} = (\theta, \max(\rho_i))$ for all $(\theta_i, \rho_i) \in S_{\theta}$.

3. \textbf{Conversion to Cartesian Coordinates} (optional for visualization):
   - Each edge point $e_{\theta}$ in polar coordinates can be converted back to Cartesian coordinates for visualization or further processing: $\text{Poly} = p_{edge} = (\rho \cos(\theta), \rho \sin(\theta))$.


\subsection{Evaluation Results}

In our experimental analysis, we conduct a comprehensive evaluation of the proposed ground segmentation method by comparing it with two existing techniques: the RANSAC method \cite{fischlerRandomSampleConsensus1981} and depth ground segmentation \cite{bogoslavskyi2017efficient}. We employ three key metrics to measure performance: the F1 score, Intersection over Union (IoU) on range images, and Bird-Eye-View (BEV) IoU that we proposed. By including evaluations based on range images and BEV areas and sequences, we endeavor to provide a comprehensive understanding of our method's performance, showcasing its effectiveness and adaptability under various conditions.

According to Table I, our method demonstrates competitive performance against state-of-the-art non-learning ground segmentation methods,i.e., RANSAC and DepthGround, as evidenced by traditional metrics such as the F1 score and IoU on range images. In terms of BEV IoU, our channel-based method, alongside DepthGround, surpasses the RANSAC approach. Remarkably, with the iteration of label propagation limited to 10, our method outperforms DepthGround, showing an approximate 12\% improvement in range-image IoU and a 30\% enhancement in BEV IoU. Throughout the sequences we evaluated, sequences 01 and 03 exhibited relatively low accuracy, primarily due to their limited size and the presence of ground truth inconsistencies, particularly in corner cases. These issues highlight the challenges in achieving high accuracy in smaller or more complex datasets, where anomalies and irregularities can significantly impact the overall performance metrics. We will delve deeper into these challenges and discuss their implications for our method's performance and reliability in Section III.E.

\begin{figure}[H]
    \includegraphics[width=\linewidth]{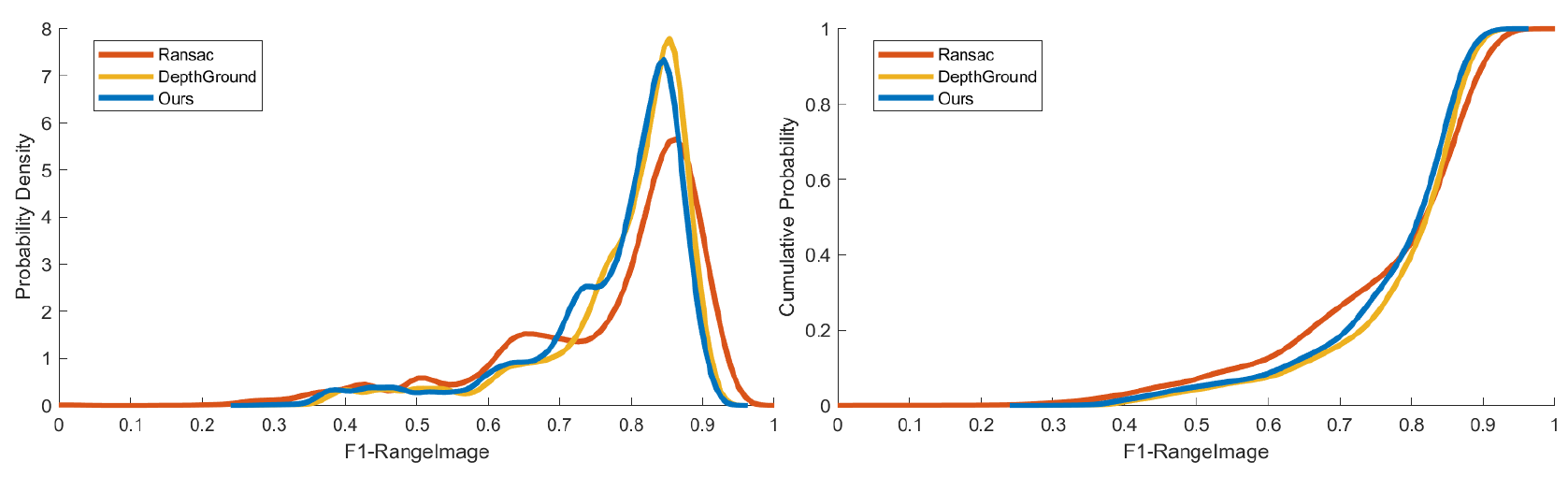}
    \caption{PDF and CDF of F1 Score Based on Range Image}
    \vspace{-0.4cm}
    \label{fig:ISICAS24_alg_perForm3}
\end{figure}

\begin{figure}[H]
    \includegraphics[width=\linewidth]{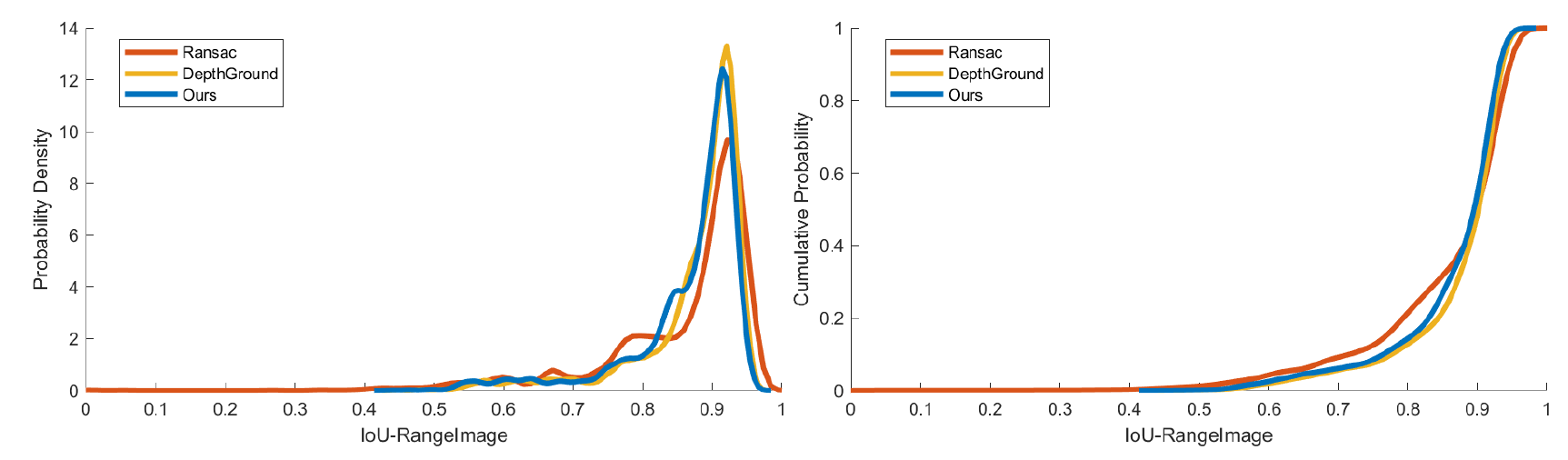}
    \caption{PDF and CDF of IoU Based on Range Image}
    \vspace{-0.4cm}
    \label{fig:ISICAS24_alg_perForm2}
\end{figure}

\begin{figure}[H]
    \includegraphics[width=\linewidth]{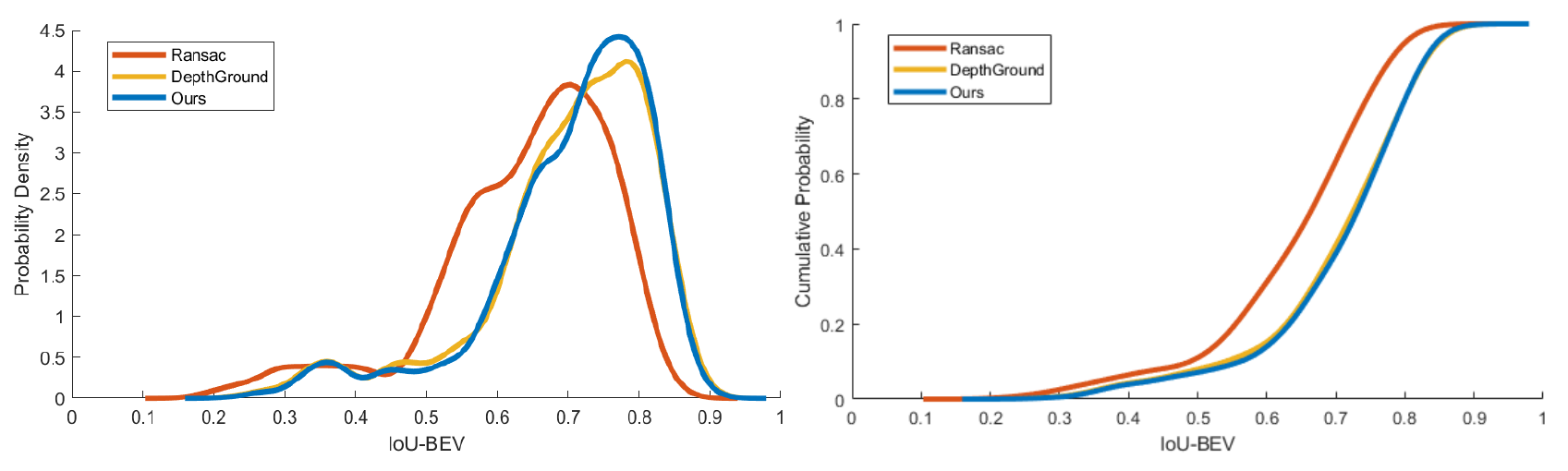}
    \caption{PDF and CDF of IoU Based on Bird-Eye-View}
    \vspace{-0.4cm}
    \label{fig:ISICAS24_alg_perForm1}
\end{figure}

For an in-depth analysis of our proposed ground segmentation method, we use sequence 00 as an illustrative example. The Probability Density Function (PDF) and Cumulative Distribution Function (CDF) of three distinct evaluation metrics are depicted in Figures \ref{fig:ISICAS24_alg_perForm3}, \ref{fig:ISICAS24_alg_perForm2}, and \ref{fig:ISICAS24_alg_perForm1}.

Figures \ref{fig:ISICAS24_alg_perForm3} and \ref{fig:ISICAS24_alg_perForm2} illustrates the advantage of our ground segmentation method, particularly highlighted through the F1 score and IoU metrics on range images. Despite the RANSAC method achieving relatively high mode values, our approach and depth ground segmentation exhibit enhanced robustness. Our approach has a denser aggregation of results within the upper echelons of quality values. On the other hand, the RANSAC method, includes performance outliers less than around 0.8 on F1 and 0.85 on IoU, indicating its variability and inconsistency across the sequence.

As previously discussed, traditional metrics like IoU-RI and F1-RI inherently emphasize the near range due to higher point density. This characteristic allows the RANSAC method to excel under these metrics by exploiting the bias. However, our proposed channel-based ground segmentation can show robustness and consistent high-quality results, even when compared to depth ground segmentation. It underscores its superior reliability and effectiveness across a broader range of conditions.

Figure \ref{fig:ISICAS24_alg_perForm3} showcases the evaluation outcomes utilizing our innovative IoU-BEV metric. Our segmentation approach outperforms the RANSAC and depth ground segmentation, marking strides in both robustness and segmentation quality. The distribution of IoU-BEV scores, reflecting the ground area coverage, presents a broader range than the tighter distributions of traditional F1-RI and IoU-RI metrics. This distinction is pivotal. Segmentation at closer ranges tends to be more straightforward due to the high data density, resulting in elevated and clustered scores with conventional metrics. By introducing an evaluation framework that focuses on the physical areas, the IoU-BEV metric accentuates the authentic segmentation strength of our method across diverse distances.

\begin{figure*} [t]
    \begin{center}
        \includegraphics[width=\linewidth]
        {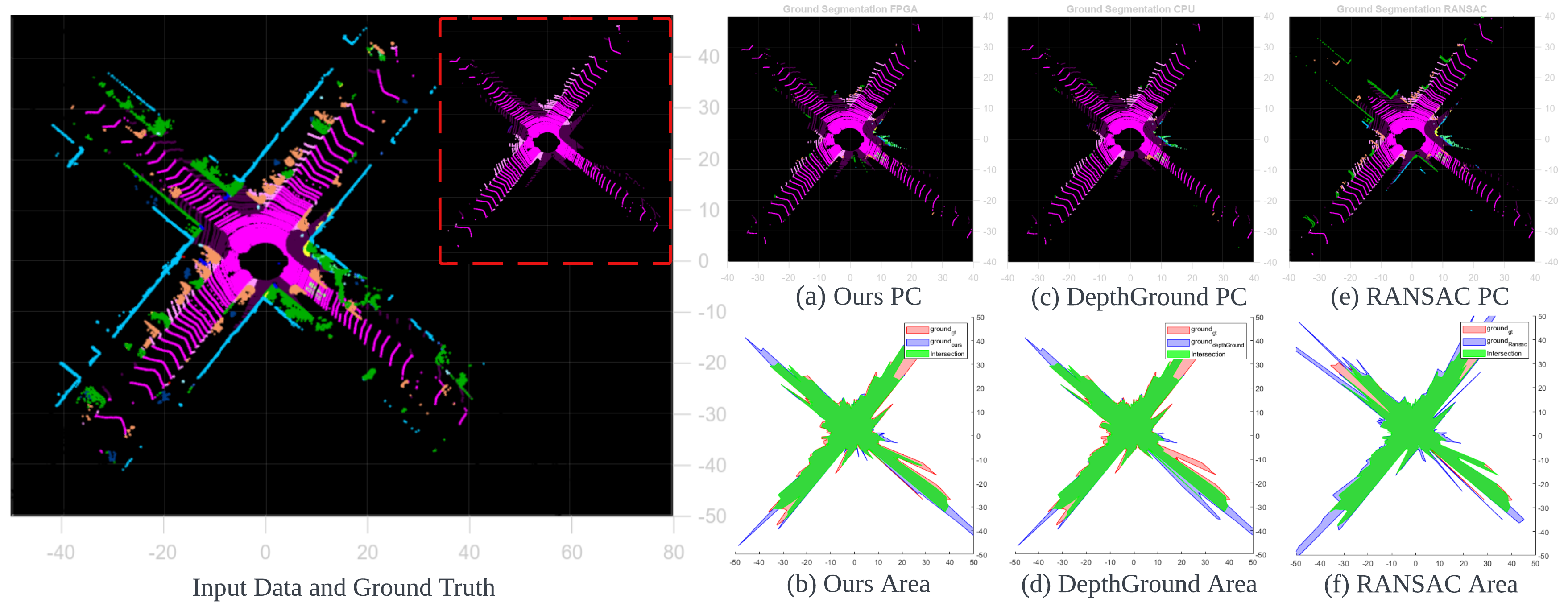}
        \caption{Case Study: Ground Segmentation in a Cross-way Scenario Using Various Methods and Evaluation Metrics. The right-hand upper sections display the point cloud outcomes  (point colors reflect semantic ground truth for enhanced result analysis). The lower sections showcase IoU-BEV area representations: green for true positives, red for false negatives, and blue for false positives. The F1 Scores for our method, depth ground segmentation, and RANSAC are 93.44, 94.25, and 95.69, respectively. IoU scores on the range image for these methods are 87.70, 0.8912, and 0.9175, respectively, while IoU scores on BEV are 0.7879, 0.7493, and 0.7228.}
        \label{fig:ISICAS24_alg_caseStudy}
    \end{center}
\end{figure*}

\subsection{Case Study}
Figure \ref{fig:ISICAS24_alg_caseStudy} presents a case study on the efficacy of different ground segmentation methods in an intersection scenario. The left section shows the input point cloud with the ground truth of ground points highlighted for clarity in the top right corner. Subsections (a) and (b) reveal the outcomes of our segmentation approach, illustrating its precision in accurately identifying the road surface. In contrast, the RANSAC method's results, depicted in subsections (e) and (f), manifest significant misclassifications, including a notable failure to detect ground in the top left road area, reflecting RANSAC's challenges with complex terrain.

While RANSAC reports the highest IoU (0.9175) and F1 scores (0.9569) on the range image, it underscores the limitations of traditional metrics in reflecting the actual ground areas. Conversely, our proposed IoU-BEV metric more accurately reflects the segmentation's effectiveness, with our method scoring the highest at 0.7879, followed by depth ground segmentation and RANSAC with scores of 0.7493 and 0.7228, respectively. This analysis underscores the IoU-BEV metric's capacity to offer a precise and realistic evaluation of ground segmentation, capturing the full scope of segmentation accuracy across the scene, even in complex environments.


\subsection{Sequence Study}

\begin{figure}[H]
    \includegraphics[width=\linewidth]{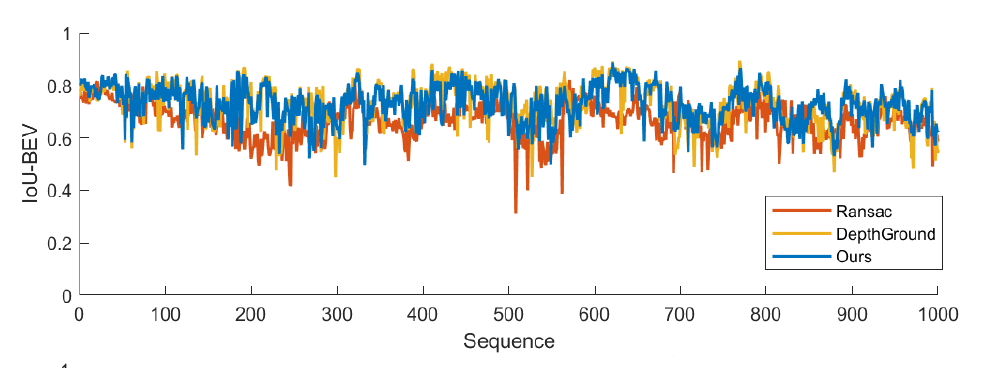}
    \caption{Comparative Performance of IoU-BEV Across 1000 Continuous Frames Among Three Different Methods}
    \vspace{-0.4cm}
    \label{fig:ISICAS24_alg_sequenceStudy1}
\end{figure}

Figure \ref{fig:ISICAS24_alg_sequenceStudy1} showcases an evaluation of ground segmentation methods across a continuous time sequence on the IoU-BEV metric. Generally, variations in scores distinctly highlight the differing performances of the methods under comparison. Notably, a score reduction can be observed between frames 500 to 550 for all methods. Detailed analysis of this segment reveals that the vehicle was navigating through a region covered with low-lying grass during this period. This scenario led to many grass patches being incorrectly classified as part of the ground plane, resulting in decreased segmentation scores.

\begin{figure}[H]
    \includegraphics[width=\linewidth]{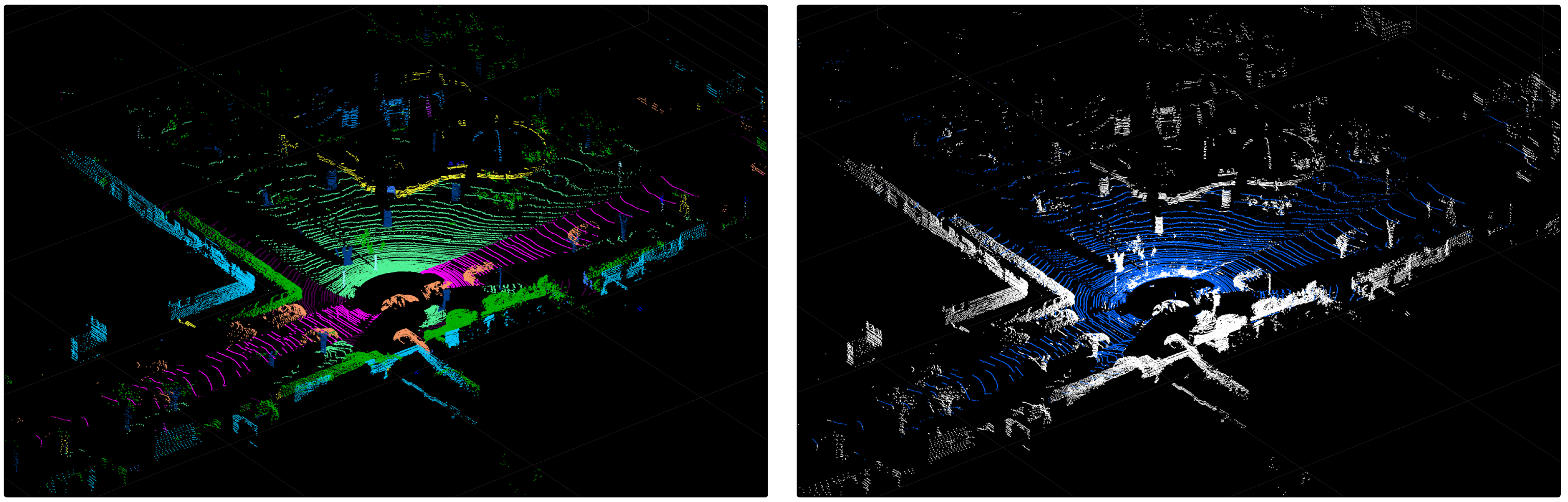}
    \caption{An Example Corner Case of Poor Evaluation Results. Left: The input point cloud with semantic labels, where green points represent vegetation class, predominantly short grass in this scenario. Right: The predicted ground segmentation outcome encompasses a large grass region. The IoU-BEV score for this frame stands at 0.5086.}
    \vspace{-0.4cm}
    \label{fig:ISICAS24_alg_sequenceStudy2}
\end{figure}

Figure \ref{fig:ISICAS24_alg_sequenceStudy2} illustrates one of these instances with the dropped score in ground segmentation. Theoretically, areas covered with short grass could be classified as part of the ground plane due to their low curvature. However, grass areas, plants, and trees are all labeled as vegetation in the SemanticKITTI dataset. This poses an additional challenge for all non-learning ground segmentation methods. This issue is not unique to the depicted sequence but a recurring problem across multiple sequences in SemanticKITTI. This example highlights the complexities of evaluating ground segmentation in real-world scenarios and established datasets.

\begin{figure*} [t]
    \begin{center}
        \includegraphics[width=13 cm]
        {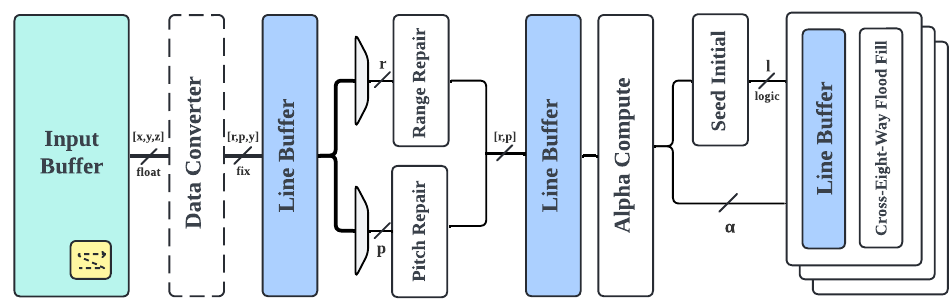}
        \caption{Illustration of the FPGA Implementation Structure: This schematic illustrates the modular architecture of the processing element, where dashed lines represent optional modules that can adapt or bypass functions based on input data.  Blue represents the line buffer, while white blocks signify combinational logic. The quantity of stacked blocks for label propagation is configurable.}
        \label{fig:ISICAS24_struct_pipeline}
    \end{center}
\end{figure*}


\section{Hardware Architecture}

\subsection{Data Path}
\begin{figure}[H]
    \includegraphics[width=\linewidth]{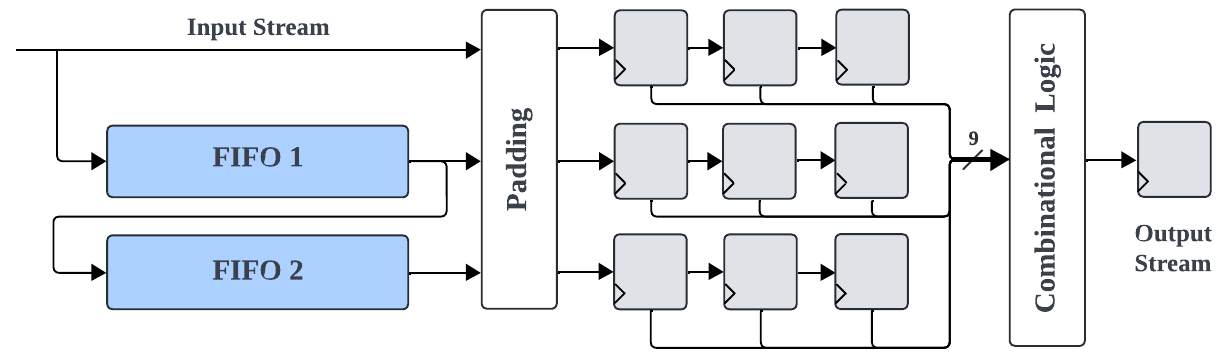}
    \caption{Line buffer Stream Processing Unit Design}
    \vspace{-0.4cm}
    \label{fig:ISICAS24_struct_linebuffer}
\end{figure}

As illustrated in Figure \ref{fig:ISICAS24_struct_pipeline}, the data path outlines the workflow of three primary stages: pre-processing, alpha matrix computation, and cross-eight-way flood-fill label propagation. Drawing inspiration from \cite{hegartyDarkroomCompilingHighlevel2014a}, we employ a line buffer pipeline throughout each phase to efficiently manage the continuous input of the point cloud. An optimized micro-architecture of the line buffer stage, depicted in Figure \ref{fig:ISICAS24_struct_linebuffer}, ensures all intermediate values within the state are preserved in a compact on-chip buffer. This approach significantly reduces both intermediate memory requirements and power consumption. As the sliced point columns from the range image transition from the input line buffer to the window select shift registers, they undergo processing by the combinational logic before advancing to the output registers. In each clock cycle, the system processes one point from input and delivers one point as output.

An additional challenge addressed during implementation, as shown in Figure \ref{fig:ISICAS24_alg_pipeline}, involves the point traversal direction in the label propagation phase, which diverges from the common image indexing way as shown in the previous phases. For the ground seed propagation, flood-fill algorithm seeds are initiated at the lowest channels of the range-image frame, leading to a search direction from bottom to top. To align with this specific requirement and ensure efficient  throughout the processing workflow, we inverted the vertical indexing of the frame to proceed from bottom-up. This modification facilitates seamless flood-fill algorithm operation, consistent with the initial seed placement, and enhances the processing pipeline's overall efficiency.

\subsection{Module Design}
\subsubsection{Data Converter}
LiDAR data typically arrives in a cartesian $[x, y, z]$ coordinate format, necessitating conversion to polar coordinates $[range, pitch, yaw]$ for our processing. A data converter is included with an optional bypass if the input data is in a polar coordinate to accommodate this.  Processing remains in the floating point to maintain accuracy.

Data conversion is followed by data quantization, where balancing precision with computational efficiency is paramount. We achieve this by converting the input data from a floating-point to a 32-bit fixed-point format. Specifically, for the range data, we utilize a fixed-point data type $\text{fixdt}(1, 32, 24)$, indicating a signed fixed-point with a total width of 32 bits—9 bits allocated for the integer part and 23 bits for the fractional part. This configuration is tailored to accommodate the typical range limits of current LiDAR data, in which the range value can be up to $2^8 - 1$ meters (255 meters). Similarly, pitch and yaw data are also represented using $\text{fixdt}(1, 32, 24)$ format, ensuring a detailed representation of angular measurements in radius.

\subsubsection{Frame Repair} The frame repair module implements the corrections of range and pitch values, employing a detailed process as previously discussed. For range value repair, we ensure smoothness in the repaired values by implementing a repair step of 5. The process begins by selecting an 11-by-1 range value array from Linebuffer's output data bus. To facilitate efficient computation of differences and averages among 5-by-5 range value pairs, we employ a matrix-based approach.

This method involves duplicating the upper (1 to 5) and lower (7 to 11) vectors of the data column to create two separate 5-by-5 matrices. Subsequent operations include matrix subtraction to determine the differences between corresponding elements and addition to prepare for averaging. Only those range pairs whose differences fall within a predefined threshold are considered valid and selected for further processing. Averaging these valid pairs is efficiently achieved through bit-shifting and reciprocal operations, ensuring a precise and smooth correction of range values.

\subsubsection{Alpha Computation} In computing alpha values, which represent the angles in the segmented ground plane, pairs of pitch and range values from adjacent channels are supplied by a 2-by-1 line buffer. The $\alpha$ calculation is then efficiently carried out using the $atan2$ module, which is implemented via the CORDIC (Coordinate Rotation Digital Computer) approximation method. This approach is chosen for its efficiency in calculating trigonometric functions, particularly suitable for FPGA implementations due to its iterative, hardware-friendly nature.

Given our bottom-up approach to frame indexing, completing $Alpha$ matrix processing allows for a straightforward final step. The top line of the alpha matrix, which may initially lack direct computational inputs due to its position, is seamlessly completed by duplicating the values from the buffered second line. This step ensures the alpha frame size consistency across the stream processing, aligning with the overall methodology designed for ground segmentation.

\subsubsection{Label Propagation} The label propagation module initiates by identifying the first valid point of each column, a process complicated by the need for zig-zag traversal. To efficiently manage this, we employ a valid seed buffer corresponding to the frame's width to track seed status. On accessing the buffer with the address of the row number, if the current valid point is within the seed threshold and the valid seed value is not yet marked as true, we label it as a seed and update the buffer to update the buffer to be true. Efficiency is further enhanced by employing a valid value flip strategy, which alternates the truth representation between '0' and '1' based on the row's parity. This method simplifies buffer initialization, allowing for quick and accurate seed initialization.

The flood-fill module takes the initial labels from the seed initialization phase and enables cross-eight-way flood-fill processing using a five-by-five line buffer. This technique is pivotal for precisely identifying and labeling ground segments within the LiDAR data. An integral part of this module is the upper channel label FIFO buffer, designed to record and update labels from upper step points efficiently. The size of this FIFO is optimized based on the algorithm's step size and frame width, ensuring efficient data flow and memory usage.

\begin{figure}[H]
    \includegraphics[width=\linewidth]{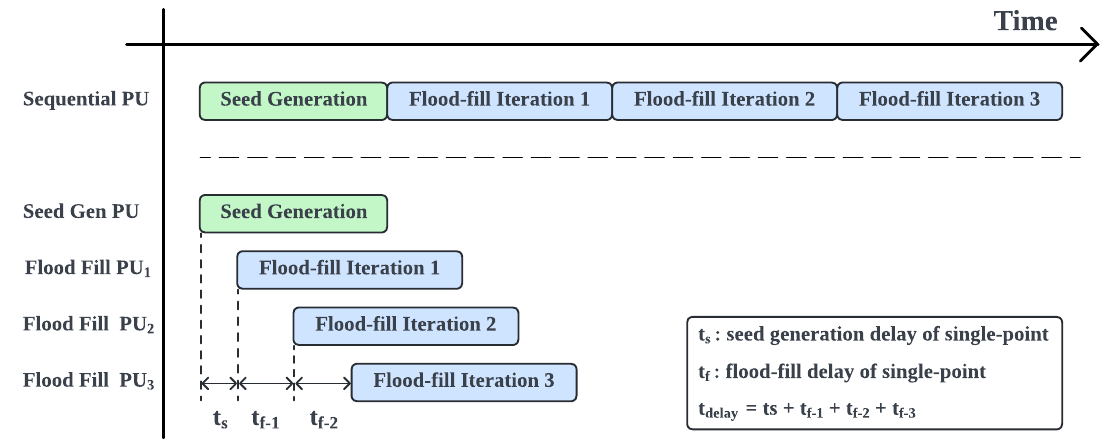}
    \caption{Schedule of label propagation}
    \vspace{-0.4cm}
    \label{fig:ISICAS24_struct_schedule}
\end{figure}

As illustrated in Figure \ref{fig:ISICAS24_struct_schedule},  pipeline architecture substantially decreases the latency compared to sequential processing, in which delay would be experienced in processing a single point results from the cumulative time taken by each phase, thereby optimizing throughput and minimizing latency by overlapping task executions. The flood-fill processing units are designed for pipelining. They are configurable to adapt to diverse performance requirements, enabling scalable deployment that can be adjusted according to desired operation speeds and resource allocation. This fixed iteration modular approach boosts computational efficiency and maintains the accuracy essential for effective ground segmentation.  By employing this strategic configuration, the module harmonizes the imperative of managing computational resources with the necessity for precise and real-time ground detection in LiDAR processing.


\subsection{Implementation Results}
Building on our previous contributions as demonstrated in the MathWorks example , we implemented the channel-based ground segmentation algorithm targeting on the Xilinx UltraScale Zynq7000 XC7Z045 FPGA. This design handles stream point cloud data pre-loaded into an on-chip buffer in range-image format, supporting various LiDAR resolutions, including 32, 64, and 128 channels (each with a horizontal resolution of 2048 in the test case).

To simulate and validate our design, we utilized HDL Coder and Vision HDL Toolbox within MATLAB 2023b, optimizing the working frequency to a robust 160 MHz. Detailed documentation of our implementation, covering resource usage, timing, and power efficiency, was generated using Xilinx Vivado 2020.1. For benchmarking, we compared our FPGA-based solution to CPU and GPU implementations. Specifically, we leveraged MATLAB's built-in functions for ground segmentation on an Intel Core i7-12700K CPU and evaluated GPU implementation outcomes on NVIDIA Geforce 560 Ti as reported by Baker and Sadowski \cite{bakerGPUAssistedProcessing2013a}.  This comprehensive approach ensures a well-rounded comparison and highlights the efficacy of our FPGA-based method.

\begin{table}[ht]
\label{tab:ISICAS24_arch_resource}
\begin{center}
\caption{Resource Usage  of channel-based ground segmentation FPGA Implementation}
\begin{tabular}{
>{\columncolor[HTML]{FFFFFF}}c |
>{\columncolor[HTML]{FFFFFF}}c |
>{\columncolor[HTML]{FFFFFF}}c |
>{\columncolor[HTML]{FFFFFF}}c }
\hline \hline
{\color[HTML]{333333} \textbf{Resource}} & {\color[HTML]{333333} \textbf{Usage}} & {\color[HTML]{333333} \textbf{Available}} & {\color[HTML]{333333} \textbf{Utilization (\%)}} \\ \hline 
Slice LUTs                               & 60395                                 & 218600                                    & 27.63                                            \\
Slice Registers                          & 76163                                 & 437200                                    & 17.42                                            \\
DSPs                                     & 26                                    & 900                                       & 2.89                                             \\
Block RAM Tile                           & 188                                   & 545                                       & 34.5         \\ \hline                                   
\end{tabular}
\end{center}

\end{table}

Since few existing FPGA implementations exist for non-learning LiDAR ground segmentation methods, direct performance comparisons with other FPGA-based designs remain elusive. However, our design showcases resource efficiency, as indicated in Table II. 
Note that our flood-fill module features three iterations as a pipeline. The resource usage suggests the potential for further optimization by increasing the iteration pipeline depth should even faster performance requirements dictate.

Timing analysis, detailed in Table III 
, demonstrates that our design comfortably supports a clock frequency of 150 MHz, with scope for further acceleration up to 160.03 MHz. The power consumption, as outlined in Table IV, is estimated at merely 2.357 W. This represents a fraction of the CPU and GPU power consumption — approximately 1.9\% of the Intel Core i7-12700K CPU's 125 Watts power usage and 1.07\% of the GeForce 560 Ti graphics card's 220 Watts \cite{bakerGPUAssistedProcessing2013a}.

\begin{table}[ht]
\begin{center}
\caption{Timing Summary of Ours FPGA Implementation}
\begin{tabular}{c|c}
\hline  \hline
\textbf{Time Constriant}           & \textbf{Values}                                     \\ \hline
\rowcolor[HTML]{FFFFFF} 
{\color[HTML]{333333} Requirement} & {\color[HTML]{333333} 6.6667 ns (150 MHz)} \\ \hline 
\rowcolor[HTML]{FFFFFF} 
Data Path Delay                    & 6.22 ns                                            \\ \hline
\rowcolor[HTML]{FFFFFF} 
Slack                              & 0.41 ns                                            \\ \hline
\rowcolor[HTML]{FFFFFF} 
Clock Frequency                    & 160.03 MHz                                          \\ \hline
\end{tabular}
\end{center}
\label{tab:ISICAS24_arch_time}
\end{table}

\begin{table}[ht]
\begin{center}
\caption{Power Estimation of Ours FPGA Implementation}
\begin{tabular}{c|c|c}
\hline \hline
\textbf{Power Type}                                                      & \textbf{Item}                                         & \textbf{Power Consumption} \\ \hline 
\cellcolor[HTML]{FFFFFF}{\color[HTML]{333333} }                          & \cellcolor[HTML]{FFFFFF}{\color[HTML]{333333} Clocks} & 0.644 W                    \\
\cellcolor[HTML]{FFFFFF}{\color[HTML]{333333} }                          & \cellcolor[HTML]{FFFFFF}Signals                       & 0.286 W                    \\
\cellcolor[HTML]{FFFFFF}{\color[HTML]{333333} }                          & \cellcolor[HTML]{FFFFFF}Logic                         & 0.290 W                    \\
\cellcolor[HTML]{FFFFFF}{\color[HTML]{333333} }                          & BRAM                                                  & 0.884 W                    \\
\multirow{-5}{*}{\cellcolor[HTML]{FFFFFF}{\color[HTML]{333333} Dynamic}} & DSP                                                   & 0.026 W                    \\ \hline
\cellcolor[HTML]{FFFFFF}Static                                           & \cellcolor[HTML]{FFFFFF}PL Static                     & 0.226 W                    \\ \hline
\end{tabular}
\end{center}
\label{tab:ISICAS24_arch_power}
\end{table}

This study of FPGA-based LiDAR ground segmentation underscores the feasibility of real-time, high-performance implementation and highlights the substantial energy savings compared to traditional computing platforms.

\subsection{Execution Time}

\begin{figure}[H]
    \includegraphics[width=\linewidth]{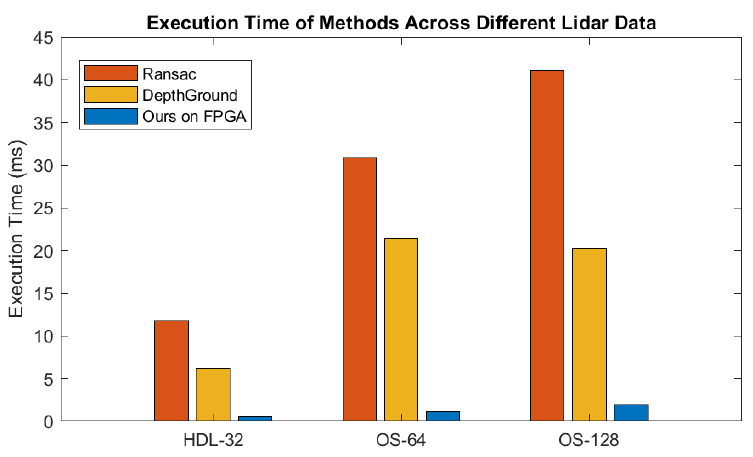}
    \caption{Execution Time (ms): A Comparison Across LiDAR Resolutions. The performance of our FPGA-based ground segmentation method demonstrates execution times of 0.54 ms, 1.09 ms, and 1.89 ms for 32-channel, 64-channel, and 128-channel LiDAR systems, respectively. These times are estimated based on the post-implementation operational frequency and simulated clock cycles.}
    \vspace{-0.4cm}
    \label{fig:ISICAS24_struct_exectime}
\end{figure}

Figure. \ref{fig:ISICAS24_struct_exectime} showcases the efficiency of our FPGA implementation tested across three LiDAR sensors: HDL-32, OS-64, and OS-128. Compared to SOTA CPU-based ground segmentation algorithms, our FPGA method significantly outperforms in speed. For the HDL-32, a low-resolution LiDAR, we observe an impressive 12-fold speed increase over CPU algorithms. With the medium-resolution OS-64 LiDAR, our design's efficiency is even more remarkable, achieving speedups of 25 and 19 times against RANSAC and depth ground segmentation methods, respectively. In handling the high-resolution OS-128 LiDAR, our approach maintains a swift processing time of 1.89ms, resulting in a speedup ranging from 10 to 22 times compared to existing SOTA methods. Therefore, our FPGA-based solution provides substantial performance improvements across different LiDAR resolutions. 

Moreover, when comparing our FPGA implementation to the GPU-based system described by Baker and Sadowski \cite{bakerGPUAssistedProcessing2013a} for HDL-64 LiDAR, our system also shows considerable advantages. With the average processing times of a dual desktop GPU setup (3.18ms) and a single GPU setup (6.33ms), our FPGA implementation achieves approximately 3 to 6 times faster processing speeds, underscoring the superior efficiency and performance of our FPGA-based approach to LiDAR ground segmentation.

\section{Conclusion}
This study introduces a channel-based LiDAR segmentation method that leverages novel angular repair techniques and a cross-eight-way flood-fill algorithm. This approach significantly reduces computational demands and hardware complexity while maintaining precise segmentation accuracy.

We demonstrate the effectiveness of the proposed method by evaluating using the SemanticKITTI dataset. In addition, we introduce an IoU-BEV metric for LiDAR point cloud segmentation. This metric provides an assessment tool for precise and practical evaluations of ground segmentation results.

A key achievement is that our optimized FPGA architecture targeted on the Zynq-7000 platform demonstrates substantial performance improvements, up to 25 times faster than the existing CPU-based implementations and 6 times faster than the GPU-based solution. In addition, our proposed hardware architecture is compatible with various LiDAR channels. This research not only proves the feasibility of low-power, real-time ground segmentation but also highlights the potential of FPGA-based solutions to autonomous vehicles and robotics technologies. 




\bibliographystyle{IEEEtran}
\bibliography{IEEEabrv,IEEEexample}

\end{document}